\documentclass[aps,twocolumn,pra,amsmath,amssymb,floatfix,showpacs]{revtex4}

\newcommand{\op}[1]{\hat{#1}}
\newcommand{\opdag}[1]{\hat{#1}^{\dag}}
\newcommand{\abs}[1]{\lvert #1 \rvert}
\newcommand{\bra}[1]{\mathinner{\langle{#1}|}}
\newcommand{\ket}[1]{\mathinner{|{#1}\rangle}}
\newcommand{\braket}[1]{\mathinner{\langle{#1}\rangle}}

\usepackage{graphicx}
\usepackage{natbib}
\usepackage{multirow}

\begin{document}

\title{Polaritonic characteristics of insulator and superfluid phases in a coupled-cavity array}

\author{E. K. Irish}
\email{e.irish@qub.ac.uk}
\affiliation{School of Mathematics and Physics, Queen's University Belfast, Belfast BT7 1NN, United Kingdom}

\author{C. D. Ogden}
\affiliation{School of Mathematics and Physics, Queen's University Belfast, Belfast BT7 1NN, United Kingdom}

\author{M. S. Kim}
\affiliation{School of Mathematics and Physics, Queen's University Belfast, Belfast BT7 1NN, United Kingdom}

\date{\today}

\begin{abstract}
Recent studies of quantum phase transitions in coupled atom-cavity arrays have focused on the similarities between such systems and the Bose-Hubbard model. However, the bipartite nature of the atom-cavity systems that make up the array introduces some differences. In order to examine the unique features of the coupled-cavity system, the behavior of a simple two-site model is studied over a wide range of parameters. Four regions are identified, in which the ground state of the system may be classified as a polaritonic insulator, a photonic superfluid, an atomic insulator, or a polaritonic superfluid.
\end{abstract}


\maketitle

\section{Introduction}

Recently a family of models for coupled arrays of atom-cavity systems has attracted considerable attention~\cite{Hartmann:2006,Greentree:2006,Angelakis:2007b,Huo:2007,Rossini:2007}. Building on the success of optical lattice experiments in demonstrating the quantum phase transition between superfluid and Mott insulator states~\cite{Greiner:2002}, these proposals have been inspired by experimental advances in photonic crystals~\cite{Armani:2003}, optical microcavities~\cite{Bayindir:2000}, and superconducting devices~\cite{Wallraff:2004}. Theoretically, they offer a fascinating combination of condensed matter physics and quantum optics. Much of the work so far has focused on the possibility of creating quantum phase transitions in systems that permit manipulation and measurement of individual lattice sites. On the practical side, several applications in quantum information processing have already been proposed. These include generation of entanglement~\cite{Angelakis:2007a}, cluster state quantum computation~\cite{Angelakis:2007c,Hartmann:2007a}, and transfer of a qubit through an array~\cite{Angelakis:2007d}.

The simplest version of the coupled-cavity model consists of a series of electromagnetic cavities, each containing a single two-level system (qubit or atom), coupled in such a way that photons may hop between adjacent cavities~\cite{Greentree:2006,Angelakis:2007b,Huo:2007}. Another model under consideration uses four-level atoms in a configuration commonly exploited for electromagnetically induced transparency, with each cavity containing multiple atoms~\cite{Hartmann:2006}. The case of several two-level atoms per cavity has been explored as well~\cite{Rossini:2007}. Studies have been carried out in the microscopic regime, with only a few cavities~\cite{Hartmann:2006,Angelakis:2007b,Huo:2007}, as well as the thermodynamic limit, in which the number of cavities goes to infinity~\cite{Greentree:2006,Rossini:2007}. 

Most of the previous work on phase transitions in coupled-cavity systems has emphasized similarities to the Bose-Hubbard model~\cite{Fisher:1989}. Indeed, the four-level-atom model can be mapped exactly onto the Bose-Hubbard Hamiltonian~\cite{Hartmann:2006}; a two-component Bose-Hubbard model has also been derived from the four-level-atom system~\cite{Hartmann:2007b}. In other coupled-cavity models, evidence has been found for a quantum phase transition between Mott insulator and superfluid states, analogous to that in the Bose-Hubbard model~\cite{Greentree:2006,Angelakis:2007b,Huo:2007,Rossini:2007,Hartmann:2007c}.

By contrast, our goal in this paper is to identify some of the unique features of the coupled-cavity system. To that end we have chosen to take a microscopic approach, building up from the well-understood Jaynes-Cummings model. Specifically, we consider a system of two cavities containing a total of two excitations. Previous work has demonstrated that many-body effects appear in finite systems of only a few cavities, including signatures of the superfluid-insulator phase transition~\cite{Hartmann:2006,Angelakis:2007b,Huo:2007}. This holds true even for a two-cavity system. (For this reason we will use some of the language of quantum phase transitions, particularly the terms ``insulator'' and ``superfluid'' for localized and delocalized states, respectively. However, it should be understood that in our usage these terms refer to states of a small finite system, not true phases in the thermodynamic sense.) The advantage of this approach is that the dimension is small enough that exact numerical solutions are easily found and, perhaps more importantly, some analytical approximations can be used. In this way we hope to identify characteristics of the coupled-cavity system that can be further explored in larger systems as well as in the thermodynamic limit.

One of the principal ways in which the coupled-cavity model differs from the Bose-Hubbard model is that two types of particles are involved. The effective on-site repulsion is provided not by a fixed classical potential but by the interaction of photons with the atom(s) in each site. This interaction depends not only on the strength of the atom-photon coupling but also on the detuning between their frequencies. The detuning, then, provides an additional parameter for the system. To further complicate the picture, the fundamental excitations in the absence of hopping are not bosons but rather entangled states of atoms and photons, known as polaritons. The relative weights of the atomic and photonic components of the polaritonic states change with the detuning. As a result, both the effective repulsion and the nature of the particles in the system depend on the detuning parameter.

In this paper we map out the parameter space of the two-cavity system over a wide range of values for the atom-cavity detuning and the photon hopping rate. The negative detuning and large hopping regimes, in particular, have  not been explored in depth previously. We identify an interesting insulator-superfluid transition in the limit of large negative detuning, which occurs at a hopping strength equal to the magnitude of the detuning. The character of this transition is distinctly different from the transition at small hopping and positive detuning studied by Angelakis et al.~\cite{Angelakis:2007b}. 

Previous studies of coupled-cavity systems have utilized the variance of the total excitation number to differentiate between insulatorlike and superfluidlike states. We also examine the atomic excitation number variance, which allows us to distinguish between purely atomic or photonic ground states and states that are composed of polaritons. Using these two measures we identify four distinct types of states: polaritonic insulator, photonic superfluid, atomic insulator, and polaritonic superfluid. In the small hopping limit, the change from atomic insulator to polaritonic insulator to photonic superfluid occurs smoothly as the detuning is increased. The situation is quite different around the transition in the regime of large negative detuning: the transition from atomic insulator to polaritonic superfluid to photonic superfluid is discontinuous to lowest order in the atom-cavity coupling parameter.  This difference highlights the complexity that can arise from the combination of individual atom-cavity dynamics and the coupling between the sites.

\section{Coupled-Cavity Model: Small Hopping Picture}\label{sec:model}

In order to keep the complexity of the system to a manageable level, we consider the simplest possible case, consisting of just two identical cavities. Each cavity supports a single field mode and contains a single two-level atom. Photons are allowed to hop between the two cavities. The Hamiltonian for the two-cavity system is given by ($\hbar = 1$)
\begin{equation}\label{hamiltonian1}
\begin{split}
H &=  \sum_{j=1,2}[\omega_c \opdag{a}_j \op{a}_j + \omega_a\ket{e_j} \bra{e_j} \\
& \quad \quad \quad + g(\opdag{a}_j \ket{g_j}\bra{e_j} + \op{a}_j \ket{e_j}\bra{g_j})] + A(\opdag{a}_1 \op{a}_2 + \opdag{a}_2 \op{a}_1) ,
\end{split}
\end{equation}
where $\omega_c$ and $\omega_a$ are the cavity and atom frequencies, respectively, $g$ is the atom-cavity coupling strength, and $A$ is the hopping strength. The operator $\op{a}_j$ ($\opdag{a}_j$) is the lowering (raising) operator for the field in cavity $j$. The states $\ket{g_j}$ and $\ket{e_j}$ represent the ground and excited states, respectively, of the atom in cavity $j$. Hence the operator $\ket{g_j}\bra{e_j}$ ($\ket{e_j}\bra{g_j}$) is the atomic lowering (raising) operator for cavity $j$.

The first term of Eq.~\eqref{hamiltonian1} gives the internal Hamiltonian for the atom-cavity systems. Individual cavities are described by the Jaynes-Cummings model, using the rotating-wave approximation~\cite{Jaynes:1963,Shore:1993}. The first two terms in the sum correspond to the internal energies of the field and atom. The interaction between the atom and the field, given by the third term, contains only so-called ``energy-conserving'' terms, in which an excitation of the field (atom) is accompanied by a deexcitation of the atom (field). Finally, the last term of Eq.~\eqref{hamiltonian1} describes the hopping of photons between the two cavities.

\begin{figure*}
\includegraphics[scale=.85]{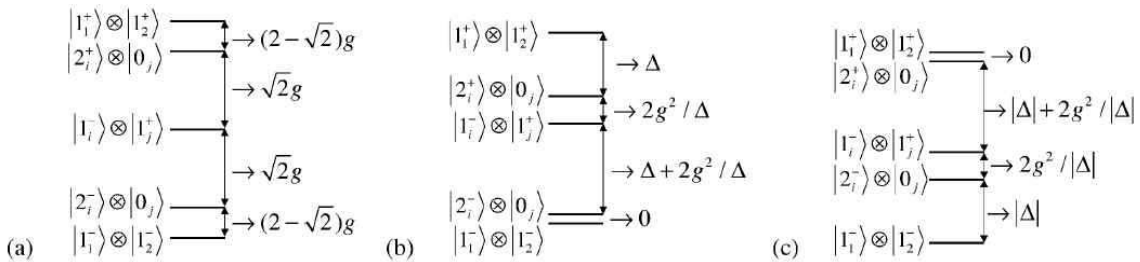}
\caption{\label{fig:energy_diagrams} Energy levels for the two-cavity system in the absence of hopping ($A=0$): (a) zero detuning ($\Delta = 0$); (b) large positive detuning ($\Delta/g \gg 1$); (c) large negative detuning ($-\Delta/g \gg 1$). }
\end{figure*}

In the absence of hopping ($A=0$) the eigenstates of the individual cavities are given by the polaritonic states
\begin{gather}
\ket{0_i} = \ket{g_i}\ket{0_i}, \label{polariton0}\\
\ket{n_i^-} = \sin \frac{\theta_n}{2} \ket{e_i}\ket{(n-1)_i} - \cos \frac{\theta_n}{2} \ket{g_i}\ket{n_i}, \label{polaritonminus} \\
\ket{n_i^+} = \cos \frac{\theta_n}{2} \ket{e_i}\ket{(n-1)_i} + \sin \frac{\theta_n}{2} \ket{g_i}\ket{n_i}, \label{polaritonplus}
\end{gather}
where $i=1,2$ denotes the cavity number, $\ket{n}$ ($n=1,2,3,\dots$) is a photon number state, and $\tan \theta_n = 2 g \sqrt{n}/\Delta$, where $\Delta = \omega_a - \omega_c$ is the detuning. The energies of these states are given by 
\begin{gather}
E_i^0 = 0 ,\\
E_i^{n\mp} = n\omega_c + \frac{\Delta}{2} \mp \frac{1}{2} \sqrt{\Delta^2 + 4ng^2}.
\end{gather}
The total number of excitations in the system $\mathcal{N} = \opdag{a}_1 \op{a}_1 + \opdag{a}_2 \op{a}_2 + \ket{e_1} \bra{e_1} + \ket{e_2} \bra{e_2}$, is conserved. In this paper the analysis is restricted to the case of exactly two excitations. There are eight possible states for the system, divided into five subspaces where states in the same subspace have degenerate energies. In order of increasing energy, the subspaces are \{$\ket{1_1^{-}}\otimes\ket{1_2^{-}}$\}, \{$\ket{2_1^{-}}\otimes\ket{0_2}$, $\ket{0_1}\otimes\ket{2_2^{-}}$\}, \{$\ket{1_1^{-}}\otimes\ket{1_2^{+}}$, $\ket{1_1^{+}}\otimes\ket{1_2^{-}}$\}, \{$\ket{2_1^{+}}\otimes\ket{0_2}$, $\ket{0_1}\otimes\ket{2_2^{+}}$\}, \{$\ket{1_1^{+}}\otimes\ket{1_2^{+}}$\}. It is important to note that the ordering of the subspaces with respect to energy is always the same regardless of the parameter values. However, the energy differences between the subspaces change significantly. Figure~\ref{fig:energy_diagrams} illustrates the energy levels in the three limiting cases of zero detuning, large positive detuning, and large negative detuning. 

\section{Total Excitation Number Variance}

Having established the model, we next turn to the problem of selecting a measure (analogous to an order parameter) that can distinguish between superfluid-like and insulator-like states. For the Bose-Hubbard model in the mean-field limit the expectation value of the boson destruction operator is typically used as the order parameter~\cite{vanOosten:2001}. In the Mott insulator state each site contains a fixed number of particles and the expectation value of the destruction operator vanishes, whereas in the superfluid state the particle number per site is not fixed and thus the expectation value becomes nonzero. However, our system is restricted to exactly two excitations. In this case the expectation value of any destruction operator is identically zero. Therefore we will first look at the ``order parameter'' utilized in Ref.~\cite{Angelakis:2007b}, which is the variance of the total excitation number in a single cavity. This quantity neatly captures the essence of the transition. In the insulator state the number of excitations per cavity is sharply defined and has zero variance. However, in the superfluid state each cavity has a finite probability of containing any number of excitations, resulting in a nonzero variance. The two cavities are, of course, completely equivalent; for definiteness we will work with cavity 1. 

\begin{figure}
\includegraphics[scale=1]{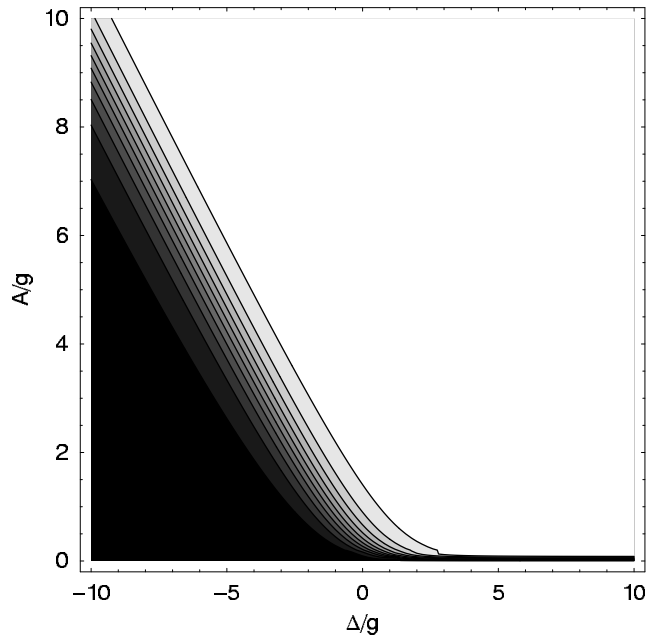}
\caption{\label{fig:varNphasediagram} Plot of $\Delta N_1$ in the ground state of the two-cavity, two-excitation system. Black corresponds to $\Delta N_1 = 0$ (Mott insulator state) while white corresponds to $\Delta N_1 = 0.5$ (superfluid state). Throughout the paper we have taken $g/\omega_a=10^{-4}$.   }
\end{figure}

The excitation number in cavity 1 and its variance are defined as, respectively,
\begin{gather}
\op{N}_1 = \opdag{a}_1 \op{a}_1 + \ket{e_1}\bra{e_1} ,\\
\Delta N_1 = \braket{\op{N}_1^2} - \braket{\op{N}_1}^2 .
\end{gather}
A contour plot of $\Delta N_1$ as a function of the detuning $\Delta$ and the hopping $A$ for the ground state of the system is shown in Fig.~\ref{fig:varNphasediagram}.

\begin{figure}
\includegraphics[scale=1]{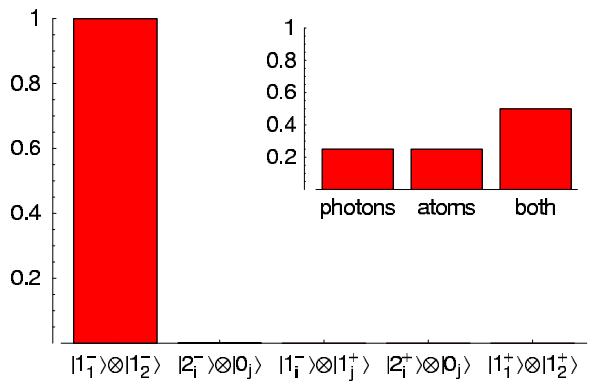}
\caption{\label{fig:polaritonic_insulator_D0}(Color online) Probability distribution, broken down by subspace, of the ground state of the system as determined by numerical diagonalization of the Hamiltonian. The inset shows the probability distribution among states with purely photonic, purely atomic, and mixed excitations. Parameter values are $\Delta = 0$, $A=g/100$. This is a Mott insulator state of polaritons. }
\end{figure}

To begin with, we consider the transition studied by Angelakis \textit{et al.}~\cite{Angelakis:2007b}, which occurs in the region defined by $\Delta \ge 0$ and $A \ll g$. At $\Delta=0$, the interaction between the atom and the field mode in a given cavity shifts the frequency of the cavity mode. This creates a photon blockade effect, prohibiting additional photons from entering the cavity~\cite{Imamoglu:1997,Birnbaum:2005}. The photon blockade leads to a large energy gap between the lowest two subspaces in the two-cavity system [see Fig.~\ref{fig:energy_diagrams}(a)]. When the hopping is weak ($A/g \ll 1$), the ground state of the system is approximately $\ket{1_1^{-}}\otimes\ket{1_2^{-}}$, as illustrated in Fig.~\ref{fig:polaritonic_insulator_D0}. Containing exactly one excitation per cavity, this state is analogous to the Mott insulator state in the Bose-Hubbard model~\cite{Fisher:1989}. 

At $\Delta = 0$ the state $\ket{1^-}$ is a fully entangled state of the atom and photon [Eq.~\eqref{polaritonminus}]. The inset of Fig.~\ref{fig:polaritonic_insulator_D0} demonstrates that, in the atom-cavity basis, this state has equal probabilities for atomic and photonic excitations. Therefore the ground state of the system at zero detuning and small hopping may be described as a polaritonic Mott insulator state.

\begin{figure}
\includegraphics[scale=1]{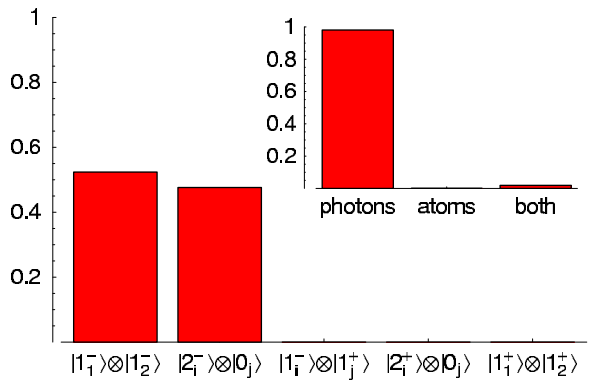}
\caption{\label{fig:large_detuning_smallA}(Color online) As Fig.~\ref{fig:polaritonic_insulator_D0}, but with $\Delta = 10g$, $A=g/100$. This is a superfluid state that is almost entirely photonic in nature.}
\end{figure}

As the detuning is increased, the energy gap becomes smaller and the photon blockade is destroyed. In the limit $\Delta/g \gg 1$ the two lowest-energy subspaces of the two-cavity system become degenerate in energy [Fig.~\ref{fig:energy_diagrams}(b)]. The lowest-order effect of the hopping is to lift the degeneracy, resulting in a unique ground state. This ground state consists of a superposition of polaritonic states such as that illustrated in Fig.~\ref{fig:large_detuning_smallA}.
However, the nature of the individual cavity eigenstates is also altered by the change in the detuning, as seen in the inset of Fig.~\ref{fig:large_detuning_smallA}. In the limit $\Delta/g \to \infty$ we have $\ket{n^-} \approx -\ket{g}\ket{n}$ and the ground state becomes $\ket{g_1}\otimes\ket{g_2}[\tfrac{1}{\sqrt{2}} \ket{1_1}\otimes \ket{1_2} - \tfrac{1}{2}(\ket{2_1}\otimes\ket{0_2} + \ket{0_1}\otimes\ket{2_2})]$. This is a delocalized photon state, i.e. a photonic superfluid. A state of this form is obtained by applying two iterations of the delocalized creation operator $\tfrac{1}{\sqrt{2}}(\opdag{a}_1 - \opdag{a}_2)$ to the vacuum state~\cite{vanOosten:2001}. This is exactly the ground state of the Bose-Hubbard model with two sites and two excitations, in the limit of large hopping.

\begin{figure}
\includegraphics[scale=1]{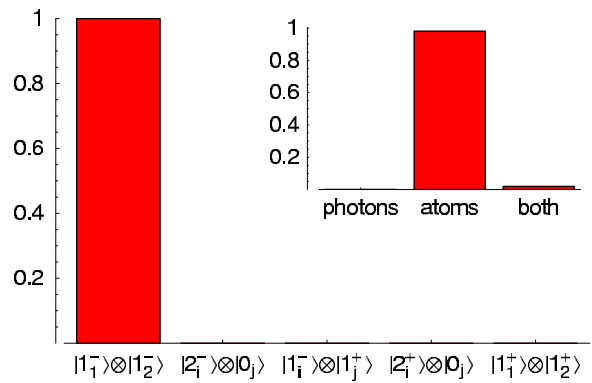}
\caption{\label{fig:atomic_insulator}(Color online) As Fig.~\ref{fig:polaritonic_insulator_D0}, but with $\Delta = -10g$, $A=g/100$. This is an insulator state composed almost entirely of atomic excitations.  }
\end{figure}

A completely different situation occurs when the detuning takes on large negative values. In this case the energy gap between the two lowest subspaces becomes larger rather than smaller, as seen in Fig.~\ref{fig:energy_diagrams}(c). The plot of $\Delta N_1$ (Fig.~\ref{fig:varNphasediagram}) indicates that the ground state of the system remains in a Mott insulator state. This state is shown in Fig.~\ref{fig:atomic_insulator}. Again, though, the nature of the atom-cavity states changes with the detuning. In the limit of large negative detuning, $\ket{1^-} \approx \ket{e}\ket{0}$ (see the inset of Fig.~\ref{fig:atomic_insulator}). Thus the ground state changes from a polaritonic insulator state at $\Delta = 0$ to an atomic insulator state at $-\Delta/g \gg 1$. The explanation for this is quite simple. From the definition of the detuning, $\Delta = \omega_a - \omega_c$, it is evident that when the detuning is negative the energy of the atoms is smaller than that of the photons. Therefore, the state of minimum energy is that in which only the atoms are excited. Because each atom is restricted to a single excitation, the purely atomic state is a localized state containing a definite number of excitations on each site, reminiscent of a Mott insulator.

\section{Atomic Excitation Number Variance}

In the analysis so far we have identified atomic and polaritonic insulator regions as well as a photonic superfluid state. A natural question, then, is whether there exists a region in which the superfluid state exhibits polaritonic characteristics. In order to answer this question, we must first define a measure that quantifies the degree to which a state is polaritonic in nature. 

Within an isolated cavity, the polariton states are characterized by a combination of atomic and photonic excitations. The total number of excitations is conserved and thus has zero variance. The variances in the numbers of atomic and photonic excitations depend on the detuning, reaching a maximum at $\Delta=0$ when the atomic and photonic degrees of freedom are maximally entangled and dropping off to zero in the large detuning limits where the states become either atomic or photonic in nature. 

In the single-cavity case, the atomic excitation number variance and the photon number variance behave similarly. However, in the coupled-cavity system, the photon number variance is nonzero in the photonic superfluid state as well as in polaritonlike states. For this reason the photon number variance is not particularly helpful for our purposes and we shall not consider it here. The atomic excitation number variance, on the other hand, is zero in both the atomic insulator state and the photonic superfluid state, and thus provides a useful measure of the polaritonic nature of the state of the system. 

\begin{figure}
\includegraphics[scale=1]{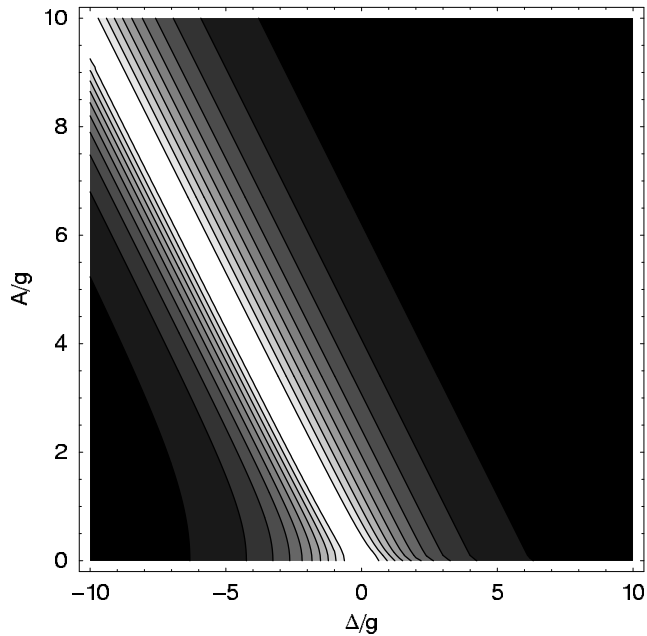}
\caption{\label{fig:varNA1AD2} Plot of $\Delta N_{A1}$ in the ground state of the two-cavity, two-excitation system. Black corresponds to $\Delta N_{A1} = 0$, while white corresponds to $\Delta N_1 = 0.25$. }
\end{figure}

A plot of the atomic excitation number variance $\Delta N_{A1}$, where $\op{N}_{A1} = \ket{e_1}\bra{e_1}$, is shown in Fig.~\ref{fig:varNA1AD2}. There are two regions in which $\Delta N_{A1}=0$. The first, with $\Delta < 0$ and $A < -\Delta$, corresponds to the atomic insulator state in which both atoms are excited. The second, which has $A > -\Delta$, corresponds to the photonic superfluid state in which both atoms are in the ground state. When $A \ll g$, the atomic excitation number variance of the ground state reduces to that of the $\ket{1^-}$ polariton in the Jaynes-Cummings model. The atomic excitation number variance peaks around $\Delta = 0$ and drops off rapidly as the magnitude of the detuning is increased. As $A$ increases, the height of the peak in $\Delta N_{A1}$ remains roughly constant, but the position of the peak follows the boundary of the insulator-superfluid transition $A \approx -\Delta$.

\begin{figure}
\includegraphics[scale=1]{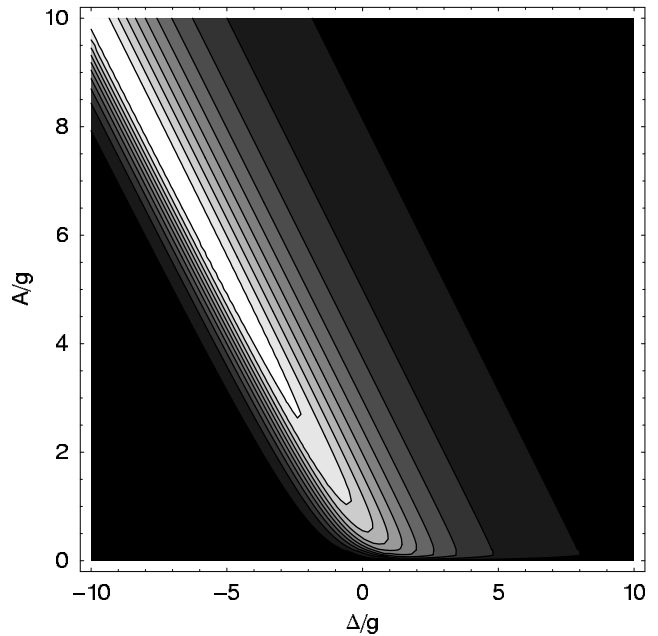}
\caption{\label{fig:variance_product} Plot of the product of the total and atomic excitation number variances $\Delta N_1 \Delta N_{A1}$ in the ground state of the two-cavity, two-excitation system. Black corresponds to $\Delta N_1 \Delta N_{A1} = 0$, while white corresponds to $\Delta N_1 \Delta N_{A1} \approx 0.1$. }
\end{figure}

The atomic excitation number variance identifies regions of polariton-like behavior, but it does not distinguish between insulator and superfluid states. In order to isolate the polaritonlike superfluid region, we take the product of the total excitation number variance $\Delta N_1$, which is nonzero in the superfluid state, and the atomic excitation number variance $\Delta N_{A1}$, which is nonzero in states with polaritonic characteristics. The result is shown in Fig.~\ref{fig:variance_product}.

\begin{figure}
\includegraphics[scale=1]{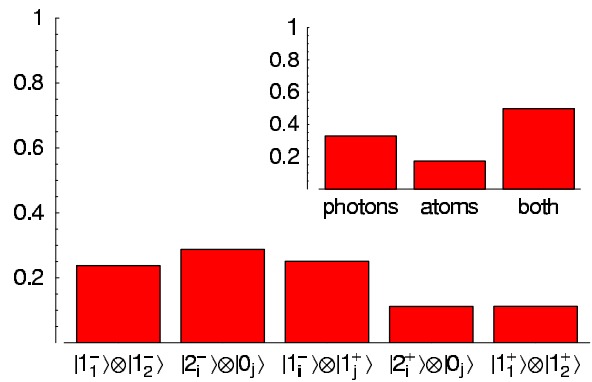}
\caption{\label{fig:polaritonic_superfluid} (Color online) As Fig.~\ref{fig:polaritonic_insulator_D0}, but with $\Delta = -10g$, $A=10g$. This represents a superfluid state with strong polaritonic characteristics. }
\end{figure}

It is apparent from Fig.~\ref{fig:variance_product} that the superfluid region does indeed overlap to some extent with the region of polaritonlike behavior. This region may be identified as a superfluid state that is, to some degree, polaritonic in nature. One such state is shown in Fig~\ref{fig:polaritonic_superfluid}. All five polariton subspaces are occupied, indicating superfluid behavior. The inset demonstrates that photonic and atomic excitations coexist, consistent with the idea that the particles involved in the superfluid state are polaritonic.

\section{Coupled-Cavity Model: Small Interaction Picture}

The approach taken in Sec.~\ref{sec:model}, in which $A$ is treated as a small parameter and the system is described in terms of polariton states, accounts for most of the behavior of the coupled-cavity system. The atomic and polaritonic insulator states and the photonic superfluid state all arise at small hopping ($A \ll g$). The shift between atomic and polaritonic behavior of the insulator state is completely determined by the atom-field interaction within each cavity, described by the Jaynes-Cummings model. In the limit of large positive detuning, the lowest-energy Jaynes-Cummings eigenstates when $A=0$ are photonic in nature, and the superfluid state arises from the zeroth-order perturbative effect of the hopping. 

The polariton-like superfluid state, on the other hand, appears only when $A \gtrsim g$. In this case the description of Sec.~\ref{sec:model} breaks down. A different approach, in which $g$ rather than $A$ is taken as the small parameter, yields greater insight into the large hopping regime and the appearance of the polaritonic superfluid. 

The Hamiltonian \eqref{hamiltonian1} may be split into three parts: a cavity Hamiltonian $H_c$, consisting of the harmonic oscillator term for each cavity plus the photon hopping term; the atomic Hamiltonian $H_a$; and the atom-cavity interaction Hamiltonian $H_i$. These are given by
\begin{align}
H_c &= \sum_{j=1,2}\omega_c \opdag{a}_j \op{a}_j + A(\opdag{a}_1 \op{a}_2 + \opdag{a}_2 \op{a}_1) ,\\
H_a &= \sum_{j=1,2}\omega_a\ket{e_j} \bra{e_j} ,\\
\begin{split}
H_i &= \sum_{j=1,2} g(\opdag{a}_j \ket{g_j}\bra{e_j} + \op{a}_j \ket{e_j}\bra{g_j}) .
\end{split}
\end{align}
Similarly, the basis states may be divided into three groups. The states that contain only photonic excitations are 
\begin{align}\label{photonic_states}
\ket{\psi_{c1}} &= \ket{g_1 1_1}\otimes\ket{g_2 1_2} ,\\
\ket{\psi_{c2}} &= \ket{g_1 2_1}\otimes\ket{g_2 0_2} ,\\
\ket{\psi_{c3}} &= \ket{g_1 0_1}\otimes\ket{g_2 2_2} .
\end{align}
Only one state contains atomic excitations alone:
\begin{equation}\label{atomic_state}
\ket{\psi_{a}} = \ket{e_1 0_1}\otimes\ket{e_2 0_2} .
\end{equation}
Finally, there are four states that each contain one photonic excitation and one atomic excitation, which are
\begin{align}\label{both_states}
\ket{\psi_{i1}} &= \ket{e_1 1_1}\otimes\ket{g_2 0_2} ,\\
\ket{\psi_{i2}} &= \ket{g_1 0_1}\otimes\ket{e_2 1_2} ,\\
\ket{\psi_{i3}} &= \ket{e_1 0_1}\otimes\ket{g_2 1_2} ,\\
\ket{\psi_{i4}} &= \ket{g_1 1_1}\otimes\ket{e_2 0_2} \label{both_last}.
\end{align}
We consider first the case $g=0$ and look for the ground state of the system. $H_c + H_a$ is block diagonal in the basis given by the states~\eqref{photonic_states}-\eqref{both_last} and may be diagonalized exactly. The resulting eigenenergies are \{$2\omega_c-2A$, $2\omega_c$, $2\omega_c+2A$, $2\omega_c+2\Delta$, $2\omega_c-A+\Delta$, $2\omega_c-A+\Delta$, $2\omega_c+A+\Delta$, $2\omega_c+A+\Delta$\}.

Noting that $A\ge0$ and $-\infty < \Delta < +\infty$, three different regimes may be identified. These regimes are distinguished by the relative values of $A$ and $\Delta$.

When $A < -\Delta$, the ground state energy is $2\omega_c - 2\Delta$, corresponding to the atomic insulator state $\ket{\psi_{a}}$. It may be seen, therefore, that the atomic insulator state is not restricted to $A \ll g$. As discussed earlier, the negative detuning provides an energy gap between the ground state and first excited state. The hopping $A$ must be on the order of $\abs{\Delta}$ in order to overcome the gap and create a superfluid state.

On the other hand, when $A > -\Delta$, the ground state energy is $2\omega_c - 2A$. This corresponds to the eigenstate $
\ket{\psi_{c1}^{\prime}} = \tfrac{1}{\sqrt{2}} \ket{\psi_{c1}} - \tfrac{1}{2}(\ket{\psi_{c2}} + \ket{\psi_{c3}})$, which is the photonic superfluid state. Evidently, the photonic superfluid state is not found only in the large positive detuning regime: it emerges when the hopping becomes large enough, regardless of the value of $\Delta$.  

\begin{figure}
\includegraphics[scale=1]{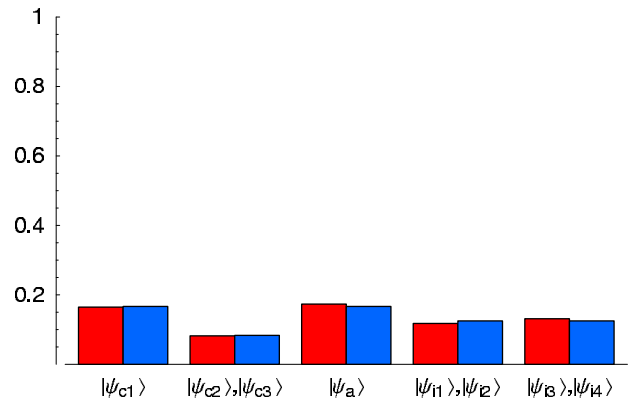}
\caption{\label{fig:degpertcompare} (Color online) Probability distribution in the atom-cavity basis of the numerically determined ground state (left-hand bars) compared with that given by Eq.~\eqref{degpertgnd} (right-hand bars). Parameter values are $\Delta = -10 g$ and $A = 10g$.}
\end{figure}

The situation becomes slightly more complicated when $A = -\Delta$. There are four degenerate eigenstates, two of which are $\ket{\psi_{c1}^{\prime}}$ and $\ket{\psi_{a}}$. The other two are given by $\ket{\psi_{i1}^{\prime}} = \tfrac{1}{\sqrt{2}}(\ket{\psi_{i2}} - \ket{\psi_{i4}})$ and $\ket{\psi_{i2}^{\prime}} = \tfrac{1}{\sqrt{2}}(\ket{\psi_{i1}} - \ket{\psi_{i3}})$. In order to identify the true ground state of the system it is necessary to take the atom-field interaction into account. Within the $4\times4$ degenerate subspace \{$\ket{\psi_{c1}^{\prime}}$,$\ket{\psi_{a}}$,$\ket{\psi_{i1}^{\prime}}$,$\ket{\psi_{i2}^{\prime}}$\}, the full Hamiltonian is given by
\begin{equation}
H = 
\begin{pmatrix} 
2\omega_c-2A & 0 & -g & -g \\
0 & 2\omega_c-2A & -\tfrac{1}{\sqrt{2}} g & -\tfrac{1}{\sqrt{2}} \\
-g & -\tfrac{1}{\sqrt{2}} & 2\omega_c-2A & 0 \\
-g & -\tfrac{1}{\sqrt{2}} & 0 & 2\omega_c-2A 
\end{pmatrix} .
\end{equation}
This matrix can be diagonalized in closed form, yielding a unique ground state with energy $E_g = 2\omega_c - 2A - \sqrt{3}g$. The corresponding eigenstate is given by
\begin{equation}\label{degpertgnd}
\ket{\phi_g} = \tfrac{1}{\sqrt{3}}\ket{\psi_{c1}^{\prime}} +  \tfrac{1}{\sqrt{6}}\ket{\psi_{a}} + \tfrac{1}{2}(\ket{\psi_{i1}^{\prime}} + \ket{\psi_{i2}^{\prime}}) .
\end{equation}
This approach constitutes a zeroth-order perturbation calculation in $g$ for the case $A = -\Delta$. The atom-cavity interaction has been included only to the extent that it lifts the ground state degeneracy. Nevertheless, the ground state \eqref{degpertgnd} provides a good approximation when $A \gg g$. A comparison with the numerical solution of the ground state in the case $A = 10g = -\Delta$ is shown in Fig.~\ref{fig:degpertcompare}. The two agree to within three percent. 

\begin{table*}
\caption{\label{tab:states}Characteristics of the four types of ground states in the coupled-cavity system.}
\begin{ruledtabular}
\begin{tabular}{ccccc}
Phase & $\Delta N_1$ & Particles & $\Delta N_{A1}$ & Regime\\ \hline
\multirow{2}{*}{Insulator } & \multirow{2}{*}{0 } & Atoms & 0 & $\Delta/g < -1, A<\abs{\Delta}$ \\
 & & Polaritons & $>0$ & $\abs{\Delta}/g \lesssim 1, A \lesssim \abs{\Delta}$ \\ \hline
\multirow{2}{*}{Superfluid } & \multirow{2}{*}{$>0$ } & Photons & 0 & $\Delta < 0, A > \abs{\Delta} \gtrsim g$; $\Delta > 0, A \gtrsim g$ \\
 & & Polaritons & $>0$ & $\Delta/g < -1, A \approx \abs{\Delta}$  \\
\end{tabular}
\end{ruledtabular}
\end{table*}

The primary lesson of this calculation is that the nature of the particles in the ground state of the coupled-cavity system is not necessarily determined by the atom-cavity interaction. The hopping term favors photonic excitations over atomic. In the positive detuning regime, photons have lower energy than atomic excitations when $A \ll g$ and so there is no competition between the interaction term and the hopping term. However, near $\Delta = 0$ the bare atom-cavity states are polaritonic. As $A$ is increased the atomic component of the ground state is gradually eliminated, leaving a purely photonic superfluid. At large negative detuning, the atom-cavity ground state is purely atomic. Near the superfluid boundary the hopping mixes the atomic and photonic components to create a polaritonlike superfluid. As the hopping is increased even further the ground state again reduces to a photonic superfluid.

\section{Conclusions}

By examining the system over a wide range of values of the hopping and detuning parameters we have uncovered some unique features of the coupled-cavity model. The four types of states we have identified are summarized in Table~\ref{tab:states}. They fall into two categories, analogous to the superfluid and insulator phases of the Bose-Hubbard model, as indicated by the total excitation number variance. Taking the atomic excitation number variance as an additional ``order parameter'' for the system allows the type of particles involved in the state to be determined. The insulator state may be either atomic or polaritonic, while the superfluid state may be photonic or polaritonic in nature.    

The different states arise from the bipartite nature of the system. Having both atoms and photons at each lattice site leads to competition between the atom-cavity interaction and the intercavity hopping. Both terms play a role in determining the phase of the system and the nature of the particles involved, although their relative importance depends on the particular parameter regime under consideration. Such richness of behavior suggests that the coupled-cavity model is more than just an analog for the Bose-Hubbard model and deserves further study in its own right.

\acknowledgments

We acknowledge support from DEL, UK EPSRC and QIPIRC.


\begin{thebibliography}{21}
\expandafter\ifx\csname natexlab\endcsname\relax\def\natexlab#1{#1}\fi
\expandafter\ifx\csname bibnamefont\endcsname\relax
  \def\bibnamefont#1{#1}\fi
\expandafter\ifx\csname bibfnamefont\endcsname\relax
  \def\bibfnamefont#1{#1}\fi
\expandafter\ifx\csname citenamefont\endcsname\relax
  \def\citenamefont#1{#1}\fi
\expandafter\ifx\csname url\endcsname\relax
  \def\url#1{\texttt{#1}}\fi
\expandafter\ifx\csname urlprefix\endcsname\relax\def\urlprefix{URL }\fi
\providecommand{\bibinfo}[2]{#2}
\providecommand{\eprint}[2][]{\url{#2}}

\bibitem[{\citenamefont{Hartmann et~al.}(2006)\citenamefont{Hartmann,
  Brand{\~a}o, and Plenio}}]{Hartmann:2006}
\bibinfo{author}{\bibfnamefont{M.~J.} \bibnamefont{Hartmann}},
  \bibinfo{author}{\bibfnamefont{F.~G. S.~L.} \bibnamefont{Brand{\~a}o}},
  \bibnamefont{and} \bibinfo{author}{\bibfnamefont{M.~B.}
  \bibnamefont{Plenio}}, \bibinfo{journal}{Nature Phys.}
  \textbf{\bibinfo{volume}{2}}, \bibinfo{pages}{849} (\bibinfo{year}{2006}).

\bibitem[{\citenamefont{Greentree et~al.}(2006)\citenamefont{Greentree, Tahan,
  Cole, and Hollenberg}}]{Greentree:2006}
\bibinfo{author}{\bibfnamefont{A.~D.} \bibnamefont{Greentree}},
  \bibinfo{author}{\bibfnamefont{C.}~\bibnamefont{Tahan}},
  \bibinfo{author}{\bibfnamefont{J.~H.} \bibnamefont{Cole}}, \bibnamefont{and}
  \bibinfo{author}{\bibfnamefont{L.~C.~L.} \bibnamefont{Hollenberg}},
  \bibinfo{journal}{Nature Phys.} \textbf{\bibinfo{volume}{2}},
  \bibinfo{pages}{856} (\bibinfo{year}{2006}).

\bibitem[{\citenamefont{Angelakis et~al.}(2007)\citenamefont{Angelakis, Santos,
  and Bose}}]{Angelakis:2007b}
\bibinfo{author}{\bibfnamefont{D.~G.} \bibnamefont{Angelakis}},
  \bibinfo{author}{\bibfnamefont{M.~F.} \bibnamefont{Santos}},
  \bibnamefont{and} \bibinfo{author}{\bibfnamefont{S.}~\bibnamefont{Bose}},
  \bibinfo{journal}{Phys. Rev. A} \textbf{\bibinfo{volume}{76}},
  \bibinfo{pages}{031805} (\bibinfo{year}{2007}).

\bibitem[{\citenamefont{Huo et~al.}()\citenamefont{Huo, Li, Song, and
  Sun}}]{Huo:2007}
\bibinfo{author}{\bibfnamefont{M.~X.} \bibnamefont{Huo}},
  \bibinfo{author}{\bibfnamefont{Y.}~\bibnamefont{Li}},
  \bibinfo{author}{\bibfnamefont{Z.}~\bibnamefont{Song}}, \bibnamefont{and}
  \bibinfo{author}{\bibfnamefont{C.~P.} \bibnamefont{Sun}},
  \bibinfo{note}{quant-ph/0702078 (2007)}.

\bibitem[{\citenamefont{Rossini and Fazio}(2007)}]{Rossini:2007}
\bibinfo{author}{\bibfnamefont{D.}~\bibnamefont{Rossini}} \bibnamefont{and}
  \bibinfo{author}{\bibfnamefont{R.}~\bibnamefont{Fazio}},
  \bibinfo{journal}{Phys. Rev. Lett.} \textbf{\bibinfo{volume}{99}},
  \bibinfo{pages}{186401} (\bibinfo{year}{2007}).

\bibitem[{\citenamefont{Greiner et~al.}(2002)\citenamefont{Greiner, Mandel,
  Esslinger, H{\"a}nsch, and Bloch}}]{Greiner:2002}
\bibinfo{author}{\bibfnamefont{M.}~\bibnamefont{Greiner}},
  \bibinfo{author}{\bibfnamefont{O.}~\bibnamefont{Mandel}},
  \bibinfo{author}{\bibfnamefont{T.}~\bibnamefont{Esslinger}},
  \bibinfo{author}{\bibfnamefont{T.~W.} \bibnamefont{H{\"a}nsch}},
  \bibnamefont{and} \bibinfo{author}{\bibfnamefont{I.}~\bibnamefont{Bloch}},
  \bibinfo{journal}{Nature (London)} \textbf{\bibinfo{volume}{415}},
  \bibinfo{pages}{39} (\bibinfo{year}{2002}).

\bibitem[{\citenamefont{Armani et~al.}(2003)\citenamefont{Armani, Kippenberg,
  Spillane, and Vahala}}]{Armani:2003}
\bibinfo{author}{\bibfnamefont{D.~K.} \bibnamefont{Armani}},
  \bibinfo{author}{\bibfnamefont{T.~J.} \bibnamefont{Kippenberg}},
  \bibinfo{author}{\bibfnamefont{S.~M.} \bibnamefont{Spillane}},
  \bibnamefont{and} \bibinfo{author}{\bibfnamefont{K.~J.}
  \bibnamefont{Vahala}}, \bibinfo{journal}{Nature (London)}
  \textbf{\bibinfo{volume}{421}}, \bibinfo{pages}{925} (\bibinfo{year}{2003}).

\bibitem[{\citenamefont{Bayindir et~al.}(2000)\citenamefont{Bayindir,
  Temelkuran, and Ozbay}}]{Bayindir:2000}
\bibinfo{author}{\bibfnamefont{M.}~\bibnamefont{Bayindir}},
  \bibinfo{author}{\bibfnamefont{B.}~\bibnamefont{Temelkuran}},
  \bibnamefont{and} \bibinfo{author}{\bibfnamefont{E.}~\bibnamefont{Ozbay}},
  \bibinfo{journal}{Phys. Rev. Lett.} \textbf{\bibinfo{volume}{84}},
  \bibinfo{pages}{2140} (\bibinfo{year}{2000}).

\bibitem[{\citenamefont{Wallraff et~al.}(2004)\citenamefont{Wallraff, Schuster,
  Blais, Frunzio, Huang, Majer, Kumar, Girvin, and Schoelkopf}}]{Wallraff:2004}
\bibinfo{author}{\bibfnamefont{A.}~\bibnamefont{Wallraff}},
  \bibinfo{author}{\bibfnamefont{D.~I.} \bibnamefont{Schuster}},
  \bibinfo{author}{\bibfnamefont{A.}~\bibnamefont{Blais}},
  \bibinfo{author}{\bibfnamefont{L.}~\bibnamefont{Frunzio}},
  \bibinfo{author}{\bibfnamefont{R.~S.} \bibnamefont{Huang}},
  \bibinfo{author}{\bibfnamefont{J.}~\bibnamefont{Majer}},
  \bibinfo{author}{\bibfnamefont{S.}~\bibnamefont{Kumar}},
  \bibinfo{author}{\bibfnamefont{S.~M.} \bibnamefont{Girvin}},
  \bibnamefont{and} \bibinfo{author}{\bibfnamefont{R.~J.}
  \bibnamefont{Schoelkopf}}, \bibinfo{journal}{Nature (London)}
  \textbf{\bibinfo{volume}{431}}, \bibinfo{pages}{162} (\bibinfo{year}{2004}).

\bibitem[{\citenamefont{Angelakis and Bose}(2007)}]{Angelakis:2007a}
\bibinfo{author}{\bibfnamefont{D.~G.} \bibnamefont{Angelakis}}
  \bibnamefont{and} \bibinfo{author}{\bibfnamefont{S.}~\bibnamefont{Bose}},
  \bibinfo{journal}{J. Opt. Soc. Am. B}
  \textbf{\bibinfo{volume}{24}}, \bibinfo{pages}{266} (\bibinfo{year}{2007}).

\bibitem[{\citenamefont{Angelakis and Kay}()}]{Angelakis:2007c}
\bibinfo{author}{\bibfnamefont{D.~G.} \bibnamefont{Angelakis}}
  \bibnamefont{and} \bibinfo{author}{\bibfnamefont{A.}~\bibnamefont{Kay}},
  \bibinfo{note}{quant-ph/0702133 (2007)}.

\bibitem[{\citenamefont{Hartmann et~al.}(2007)\citenamefont{Hartmann, Brand{\~a}o, and Plenio}}]{Hartmann:2007a}
\bibinfo{author}{\bibfnamefont{M.~J.} \bibnamefont{Hartmann}},
  \bibinfo{author}{\bibfnamefont{F.~G. S.~L.} \bibnamefont{Brand{\~a}o}},
  \bibnamefont{and} \bibinfo{author}{\bibfnamefont{M.~B.}
  \bibnamefont{Plenio}}, \bibinfo{journal}{Phys. Rev. Lett.}
  \textbf{\bibinfo{volume}{99}}, \bibinfo{pages}{160501}
  (\bibinfo{year}{2007}).

\bibitem[{\citenamefont{Bose et~al.}(2007)\citenamefont{Bose, Angelakis, and
  Burgarth}}]{Angelakis:2007d}
\bibinfo{author}{\bibfnamefont{S.}~\bibnamefont{Bose}},
  \bibinfo{author}{\bibfnamefont{D.~G.} \bibnamefont{Angelakis}},
  \bibnamefont{and} \bibinfo{author}{\bibfnamefont{D.}~\bibnamefont{Burgarth}},
  \bibinfo{journal}{J. Mod. Opt.} \textbf{\bibinfo{volume}{54}},
  \bibinfo{pages}{2307} (\bibinfo{year}{2007}).

\bibitem[{\citenamefont{Fisher et~al.}(1989)\citenamefont{Fisher, Weichman,
  Grinstein, and Fisher}}]{Fisher:1989}
\bibinfo{author}{\bibfnamefont{M.~P.~A.} \bibnamefont{Fisher}},
  \bibinfo{author}{\bibfnamefont{P.~B.} \bibnamefont{Weichman}},
  \bibinfo{author}{\bibfnamefont{G.}~\bibnamefont{Grinstein}},
  \bibnamefont{and} \bibinfo{author}{\bibfnamefont{D.~S.}
  \bibnamefont{Fisher}}, \bibinfo{journal}{Phys. Rev. B}
  \textbf{\bibinfo{volume}{40}}, \bibinfo{pages}{546} (\bibinfo{year}{1989}).

\bibitem[{\citenamefont{Hartmann et~al.}()\citenamefont{Hartmann, Brand{\~a}o,
  and Plenio}}]{Hartmann:2007b}
\bibinfo{author}{\bibfnamefont{M.~J.} \bibnamefont{Hartmann}},
  \bibinfo{author}{\bibfnamefont{F.~G. S.~L.} \bibnamefont{Brand{\~a}o}},
  \bibnamefont{and} \bibinfo{author}{\bibfnamefont{M.~B.}
  \bibnamefont{Plenio}}, \bibinfo{note}{arXiv:0706.2251 (2007)}.

\bibitem[{\citenamefont{Hartmann and Plenio}(2007)}]{Hartmann:2007c}
\bibinfo{author}{\bibfnamefont{M.~J.} \bibnamefont{Hartmann}} \bibnamefont{and}
  \bibinfo{author}{\bibfnamefont{M.~B.} \bibnamefont{Plenio}},
  \bibinfo{journal}{Phys. Rev. Lett.} \textbf{\bibinfo{volume}{99}},
  \bibinfo{pages}{103601} (\bibinfo{year}{2007}).

\bibitem[{\citenamefont{Jaynes and Cummings}(1963)}]{Jaynes:1963}
\bibinfo{author}{\bibfnamefont{E.~T.} \bibnamefont{Jaynes}} \bibnamefont{and}
  \bibinfo{author}{\bibfnamefont{F.~W.} \bibnamefont{Cummings}},
  \bibinfo{journal}{Proc. IEEE} \textbf{\bibinfo{volume}{51}},
  \bibinfo{pages}{89} (\bibinfo{year}{1963}).

\bibitem[{\citenamefont{Shore and Knight}(1993)}]{Shore:1993}
\bibinfo{author}{\bibfnamefont{B.~W.} \bibnamefont{Shore}} \bibnamefont{and}
  \bibinfo{author}{\bibfnamefont{P.~L.} \bibnamefont{Knight}},
  \bibinfo{journal}{J. Mod. Optics} \textbf{\bibinfo{volume}{40}},
  \bibinfo{pages}{1195} (\bibinfo{year}{1993}).

\bibitem[{\citenamefont{van Oosten et~al.}(2001)\citenamefont{van Oosten,
  van~der Straten, and Stoof}}]{vanOosten:2001}
\bibinfo{author}{\bibfnamefont{D.}~\bibnamefont{van Oosten}},
  \bibinfo{author}{\bibfnamefont{P.}~\bibnamefont{van~der Straten}},
  \bibnamefont{and} \bibinfo{author}{\bibfnamefont{H.~T.~C.}
  \bibnamefont{Stoof}}, \bibinfo{journal}{Phys. Rev. A}
  \textbf{\bibinfo{volume}{63}}, \bibinfo{pages}{053601}
  (\bibinfo{year}{2001}).

\bibitem[{\citenamefont{Imamo{\=g}lu et~al.}(1997)\citenamefont{Imamo{\=g}lu,
  Schmidt, Woods, and Deutsch}}]{Imamoglu:1997}
\bibinfo{author}{\bibfnamefont{A.}~\bibnamefont{Imamo{\=g}lu}},
  \bibinfo{author}{\bibfnamefont{H.}~\bibnamefont{Schmidt}},
  \bibinfo{author}{\bibfnamefont{G.}~\bibnamefont{Woods}}, \bibnamefont{and}
  \bibinfo{author}{\bibfnamefont{M.}~\bibnamefont{Deutsch}},
  \bibinfo{journal}{Phys. Rev. Lett.} \textbf{\bibinfo{volume}{79}},
  \bibinfo{pages}{1467} (\bibinfo{year}{1997}).

\bibitem[{\citenamefont{Birnbaum et~al.}(2005)\citenamefont{Birnbaum, Boca,
  Miller, Boozer, Northup, and Kimble}}]{Birnbaum:2005}
\bibinfo{author}{\bibfnamefont{K.~M.} \bibnamefont{Birnbaum}},
  \bibinfo{author}{\bibfnamefont{A.}~\bibnamefont{Boca}},
  \bibinfo{author}{\bibfnamefont{R.}~\bibnamefont{Miller}},
  \bibinfo{author}{\bibfnamefont{A.~D.} \bibnamefont{Boozer}},
  \bibinfo{author}{\bibfnamefont{T.~E.} \bibnamefont{Northup}},
  \bibnamefont{and} \bibinfo{author}{\bibfnamefont{H.~J.}
  \bibnamefont{Kimble}}, \bibinfo{journal}{Nature (London)}
  \textbf{\bibinfo{volume}{436}}, \bibinfo{pages}{87} (\bibinfo{year}{2005}).

\end{thebibliography}
\end{document}